\documentclass[11pt, tightenlines, onecolumn, amsmath, amssymb, notitlepage, a4paper, nofootinbib, superscriptaddress]{revtex4-2}

\usepackage[strict]{revquantum}
\usepackage{graphicx}
\usepackage{braket}
\usepackage[margin=2cm]{geometry}
\usepackage{multirow}
\usepackage{tabularx}
\usepackage{array}
\usepackage[normalem]{ulem}
\usepackage{fancyhdr}

\DeclareMathOperator{\tr}{tr}

\algnewcommand\algin{\textbf{in}}

\definecolor{forestgreen}{rgb}{0.13, 0.55, 0.13}

\makeatletter
\let\ps@titlepage\ps@plain
\makeatother

\begin{document}

\title{Efficient Quantum State Tracking in Noisy Environments}

\author{Markus Rambach}
\email{m.rambach@uq.edu.au}
\affiliation{Australian Research Council Centre of Excellence for Engineered Quantum Systems \& School of Mathematics and Physics, University of Queensland, QLD 4072, Australia.}

\author{Akram Youssry}
\affiliation{University of Technology Sydney, Centre for Quantum Software and Information, Ultimo NSW 2007, Australia.}
\affiliation{Quantum Photonics Laboratory and Centre for Quantum Computation and Communication Technology, RMIT University, Melbourne, VIC 3000, Australia} 

\author{Marco Tomamichel}
\affiliation{Department of Electrical and Computer Engineering \& Centre for Quantum Technologies, National University of Singapore, Singapore 119077, Singapore.}

\author{Jacquiline Romero}
\affiliation{Australian Research Council Centre of Excellence for Engineered Quantum Systems \& School of Mathematics and Physics, University of Queensland, QLD 4072, Australia.}

\setcounter{secnumdepth}{0}

\begin{abstract}

\noindent
Quantum state tomography, which aims to find the best description of a quantum state---the density matrix, is an essential building block in quantum computation and communication. 
Standard techniques for state tomography are incapable of tracking changing states and often perform poorly in the presence of environmental noise. 
Although there are different approaches to solve these problems theoretically, experimental demonstrations have so far been sparse. 
Our approach, matrix-exponentiated gradient tomography, is an online tomography method that allows for state tracking, updates the estimated density matrix dynamically from the very first measurements, is computationally efficient, and converges to a good estimate quickly even with noisy data. 
The algorithm is controlled via a single parameter, its learning rate, which determines the performance and can be tailored in simulations to the individual experiment. 
We present an experimental implementation of matrix-exponentiated gradient tomography on a qutrit system encoded in the transverse spatial mode of photons.
We investigate the performance of our method on stationary and evolving states, as well as significant environmental noise, and find fidelities of around 95\% in all cases.

\end{abstract}


\maketitle

\section{Introduction}

\noindent
Characterising quantum systems becomes increasingly important as quantum technologies begin to scale up. 
Experiments often require verification of the prepared quantum state and detection of errors as the state evolves, e.g., through deliberate evolution or environmental perturbations. 
Thus, quantum state tomography---finding the density matrix that best describes a quantum system---is a task central to quantum information processing. 
However, quantum state tomography (QST) is notoriously resource-intensive~\cite{Paris2004}. 
Reconstructing the $d$-by-$d$ density matrix of a $d$-dimensional quantum system up to a fixed precision requires $O(d^2)$ parameters that can only be obtained after measuring at least $O(d^3)$ copies of the quantum state~\cite{Haah17}.
Most common algorithms for QST (e.g. maximum likelihood estimation~\cite{Dariano2003} and least squares regression~\cite{Opatrny1997}) require a tomographically complete set of measurements and are done in post-processing, thus precluding real-time control of experiments based on the QST results. 

Current experimental realisations of quantum devices often suffer from perturbations that change over time which might happen faster than a complete set of measurements can be performed. 
It might also be of interest to the experimentalist to observe the evolution of states in an experiment.
Both these scenarios require online QST---learning algorithms that continuously update the estimate of the state description~\cite{Aaronson2019}, and thus enable real-time control and diagnosis of errors. 
Recent proposals for this include Bayesian approaches~\cite{Granade2016,Lin2020}, adaptive measurements~\cite{Qi2017a,Chen2020,Nohara2020,Quek2021a}, and machine learning techniques~\cite{Ferrie2014,Li2019,Youssry2019,Zhang2020,Quek2021a}. 
All these techniques are also closely related to continuous learning~\cite{Silberfarb2005,Shabani2014}---weak measurements over time to characterise a systems evolution, motivated by feedback control to correct errors---and Hamiltonian identification/learning~\cite{Cole2005,Anshu2021}---algorithms to determine Hamiltonian parameters governing the dynamics or unknown structures in the system, motivated by distinguishing and quantification of errors.

Moreover, arbitrary noise from the environment or from imperfect measurements can easily surpass the signal strength in experiments.
This degrades our ability to accurately estimate the quantum state, hence the robustness of techniques is crucial.
Usual efforts to circumvent this issue experimentally try to improve the signal-to-noise ratio (SNR), which is not always feasible or indeed possible.
Some theoretical work exist, e.g., Ref.~\cite{Bogdanov2016a,Ivanova-Rohling2022} where the noise is formalised into the measurement, or  Ref.~\cite{Farooq2022} where an algorithm robust to noisy data is designed.
However, none of these consider dynamical states and we are not aware of any experimental demonstrations.

Here, we focus on quantum state tomography using the matrix-exponentiated gradient method (MEG)~\cite{Li2019,Youssry2019}. 
The reduced computational complexity per iteration of MEG, $O(d^3)$ in contrast to maximum likelihood and least squares regression which are both $O(d^4)$, makes it a good candidate for online quantum state tomography. 
The MEG algorithm is also efficient---it does not require a full set of measurements for each state estimate---and robust---simulations show convergence even with very noisy measurements. 
Additionally, the MEG update rule ensures that the estimated state is physical (i.e., a positive-semi definite density matrix), which is generally not guaranteed in most tomography methods unless the estimate is projected back into the physical space, causing a bias towards low-rank states in the estimator~\cite{Blume-Kohout2010}.

In this work, we experimentally demonstrate real-time, online quantum state tomography based on MEG using a high-dimensional quantum system encoded in the transverse spatial mode of single photons. 
We achieve fidelities of up to 95\% even in the presence of significant noise due to statistical fluctuations (e.g. very low count rates) and environmental effects (e.g. an ambient light source).
To our knowledge, this is the first experimental $O(d^3)$ online tomography implementation to date.
A similar $O(d^3)$ online algorithm has been proposed recently~\cite{Zhang2020}, however, an experiment is still lacking.

\section{Results}
\subsection{Overview} 

\noindent 
Matrix-exponentiated gradient tomography (MEG)~\cite{Li2019,Youssry2019} is an online algorithm that can estimate and track quantum states, adapted from machine learning techniques. 
The idea is to construct an iterative procedure, in which an estimate $\hat{\rho}_t$ of the (mixed or pure) quantum state at the iteration $t$ is updated to a more accurate estimate $\hat{\rho}_{t+1}$, given a new measurement performed on the unknown quantum state. 
The matrix-exponentiated gradient update rule~\cite{Youssry2019} for the estimate $\hat{\rho}_{t+1}$ is given by
\begin{align}
    \hat{\rho}_{t+1} = \frac{\exp(\log(\hat{\rho_t}) - \eta_t\nabla L_t)}{\tr{\exp(\log(\hat{\rho_t}) - \eta_t\nabla L_t)}},
\end{align}
where $\eta_t$ is the learning rate, and $L_t$ is the loss function,
\begin{align}
    L_t = (\tr(\hat{\rho}_{t}X_t) - y_t)^2,
\end{align}
and its gradient is given by
\begin{align}
    \nabla L_t = 2(\tr(\hat{\rho}_{t}X_t) - y_t)X_t.
\end{align}
The pair $(X_t, y_t)$ denotes the measurement record where $X_t$ is the measurement operator and $y_t$ is the experimentally obtained average value of measuring the state with operator $X_t$. 
The measurements have to be informationally-complete as a requirement for performing any quantum tomography procedure. 
Therefore, at each iteration, a measurement operator $X_t$ is chosen at random from a complete set of bases, and together with the measurement result $y_t$, the estimate of the state $\hat{\rho}_{t}$ is updated to $\hat{\rho}_{t+1}$. 
The learning rate $\eta_t$ determines the weight given to the new obtained information at each iteration.
For tomography on stationary states, $\eta_t$ can be chosen to decrease continuously with the number of iterations while for online state tracking it needs to be constant or adaptive, depending on former measurement outcomes.

Compared to the originally proposed MEG in Ref.~\cite{Youssry2019}, we made a few changes and extensions to reflect the conditions in our experiment (more details in the METHODS section).
First, we extended MEG to work with general higher-dimensional systems $d>2$, beyond qubit systems and local Pauli measurements, by measuring mutually unbiased bases (MUBs)~\cite{Durt2010}. However, not all MUBs are known for arbitrary dimensions so we introduced a second scheme which uses generalised Pauli operator measurements~\cite{Kimura2003}.
These operators form an orthonormal informationally-complete basis set and can be used to expand MEG to arbitrary high dimensions.
Second, we modified the update rule in order to make better use of the experimental data.
MEG originally drew on the probability of one measured state per iteration to update its estimate.
Experimentally, we measure photon counts and need the counts of all $d$ states of a given basis to calculate probabilities.
This means that in each iteration we can calculate $d$ probabilities which are now all employed to update the estimate.
Third, we also adapted MEG to account for the fact that the prepared states in our experiment are very close to pure states---true for systems that have very high control over their preparation. 
In each iteration $t$, we find the eigenvalues and eigenvectors of the estimated density matrix $\hat{\rho}_t$ and pick the eigenvector with the largest eigenvalue as our estimated state $\ket{\phi_t}$ (density matrix $\hat{\rho}_{\phi}$) for benchmarking.
We then calculate the infidelity $1 - f(\hat{\rho}_{\phi},\hat{\Omega}_{\psi}) = 1 - \left( \mathrm{tr} \sqrt{\sqrt{\hat{\rho}_{\phi}} \, \hat{\Omega}_{\psi} \, \sqrt{\hat{\rho}_{\phi}}}  \right)^2$, 
where $f(\hat{\rho}_{\phi},\hat{\Omega}_{\psi})$ is the fidelity and $\hat{\Omega}_{\psi}$ is the density matrix of the theoretical prepared state $\ket{\psi_t}$. 
Our results show convergence to the unknown state within experimental limitations, most prominently mode-dependent losses, in all investigated situations and even for excessive experimental noise.

\subsection{State Estimation and Online Tracking}
\noindent 
We investigated the qutrit ($d\,=\,3$) state estimation and online tracking performance of MEG using mutually unbiased bases (MUBs) and generalised Pauli operator measurements, which makes the algorithm applicable to higher dimensions. 
We also investigated MEG performance in the presence of small and large statistical and environmental noise.
In all cases we continuously evolve the prepared state in time following
\begin{equation}
    \label{eqn:state_evo}
    \ket{\psi_t} = \exp{\left( -i \sigma \omega t \right)} \ket{\psi_0},
\end{equation}
with $\sigma$ a Hermitian matrix and $\omega = \tfrac{1.3}{t_{\text{tot}}}$ the rate of change dependent on the total amount of iterations $t_{\text{tot}}$.
The rate of change was chosen to allow the state to evolve to a minimum fidelity compared to the initial state and then back to its initial state.
We studied three possible cases of $\sigma$: $0_{3,3}$ which is the 3-dimensional zero matrix, $\sigma_z$ which is the generalised 3-dimensional \textit{Pauli Z} matrix given by
\begin{align}
    \sigma_z = \frac{1}{\sqrt{3}} \begin{pmatrix}1 & 0& 0 \\ 0 & 1 & 0\\ 0 & 0 & -2 \end{pmatrix},
\end{align}  
and $\sigma_r$ which is a general random Hermitian matrix. 
The zero matrix results in the identity evolution leaving the initial state stationary, while the other two matrices simulate a rotation in the Hilbert space of a qutrit. 
To ensure a fair comparison, the learning rate---a crucial parameter for performance and convergence of the algorithm---is kept constant at $\eta = 5$, which shows good performance in all cases albeit not optimal for $\sigma = 0_{3,3}$.
The learning rate can be optimised individually, dependent on investigated scenario and physical system. 

\begin{figure}[t]
\includegraphics[width=\columnwidth]{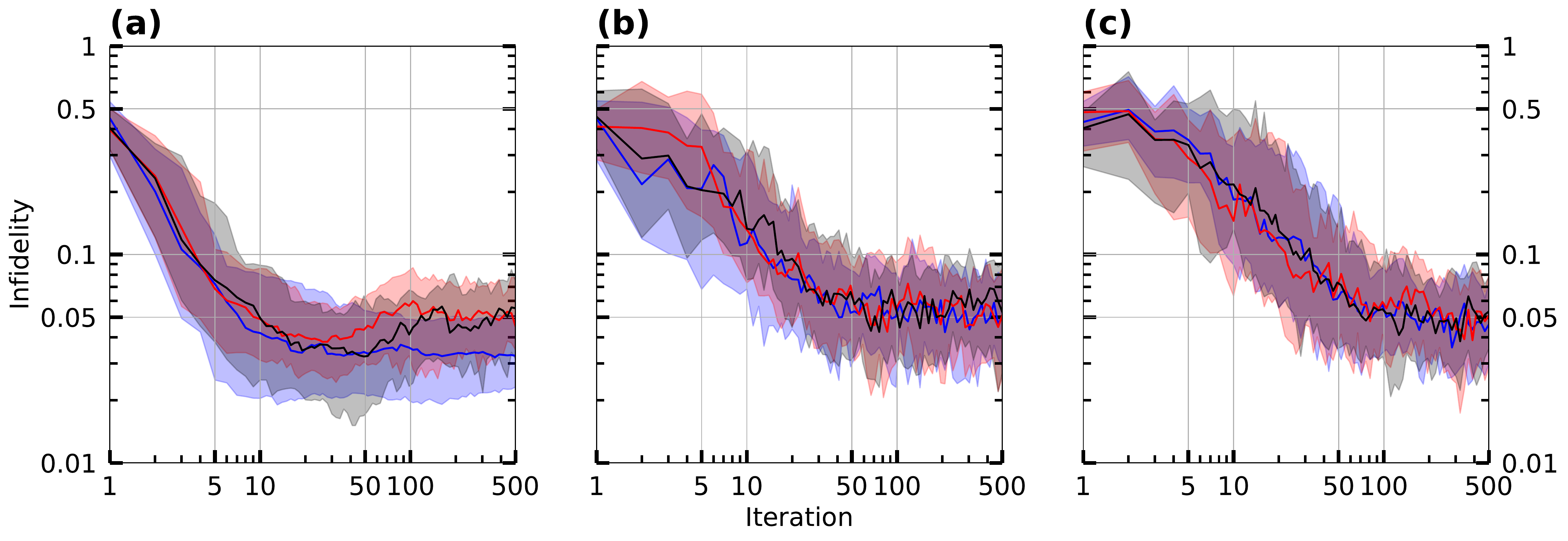}
\caption{
\label{fig:MEG_all}
MEG on qutrits for (a) high signal rates ($N \approx 10^6$ photons per measurement) using MUBs, (b) low signal rates ($N \approx 10^2$ photons per measurement) using MUBs, and (c) low signal rates using generalised Pauli operators.
In each case, three scenarios according to different time evolutions of the states (Eq.~\ref{eqn:state_evo}) are investigated: stationary (blue), $\sigma_z$ (red), and a random unitary $\sigma_r$ (black).
Solid lines are median performances of the algorithm out of 50 randomly chosen states (according to Haar measure).
The shaded regions are the interquartile ranges.
}
\end{figure}

An initial baseline measurement using MUBs and high signal rates ($N \approx 10^6$ photons per measurement) is shown in Fig.~\ref{fig:MEG_all}(a). 
The high count rate ensures that statistical Poissonian distributed counting noise ($\Delta N = \sqrt{N} \approx 10^3$), background ($N_{\mathrm{back}} \approx 50$) and dark counts ($N_{\mathrm{dark}} \approx 100$), are negligible and do not deteriorate the performance.
For this case, the algorithm finds density matrices with a median purity above 99.6\% in all investigated scenarios, as to be expected for our system where we have a high degree of control over the state preparation.
Since the fidelities in the experiments are very high, we plot the infidelity, $1 - f(\hat{\rho}_{\phi},\hat{\Omega}_{\psi})$, for clarity.
The solid lines in Fig.~\ref{fig:MEG_all}(a) are the median infidelities of 50 randomly chosen states of qutrits (according to Haar measure), with the shaded regions bounded by the upper and lower quartile $(50 \pm 25)\%$.

In theory, MEG can estimate any quantum state with arbitrary precision and accuracy given an infinite amount of iterations.
However, experimental imperfections such as misalignment and mode-dependent loss significantly limit the minimum infidelities achievable by QST algorithms (not just MEG). 
The infidelities for MEG listed in Tab.~\ref{tab:values} are similar to those obtained using other quantum state tomography algorithms applied to our system, e.g. root-approach and maximum-likelihood estimators~\cite{Bogdanov2009} where we set set the rank to 1. 

We next show the robustness of MEG to noise using MUBs, reducing the count rate per iteration to $N \approx 10^2$.
This means that the values for Poissonian, dark, and background noise become comparable and play a significant role in state estimation and tracking.
Robustness to noise in the limit of a small number of repeated preparations of the unknown state---here the number of photons---is especially interesting for systems like ions and superconducting qubits, as the preparation is considerably more time consuming compared to our system.
Again, we benchmark the performance of MEG by calculating the infidelity in each iteration, see Fig.~\ref{fig:MEG_all}(b). 
State estimation and online tracking are in excellent agreement with an overall mean infidelity $\sim~5.5\%$, slightly higher than in the high count rate case, but well within the uncertainty bars.
This already shows the robustness of the modified algorithm to noise.
Infidelities below $10\%$ are achieved within around 15 iterations, four times more than in the baseline measurement with high count rates (see Tab.~\ref{tab:values} for details).

We furthermore studied the performance of MEG under high noise, but using generalised Pauli operator measurements rather than MUBs, shown in Fig.~\ref{fig:MEG_all}(c). 
As generalised Pauli operators can be mathematically defined using them instead of MUBs is a powerful tool to describe states far beyond the qubit.
We observe that state estimation and online tracking are again in excellent agreement and the overall mean infidelity $\sim~5.1\%$ is slightly better than in the MUB case, but overlapping within uncertainty bars.
This demonstrates that generalised Pauli measurements is a viable path to extend MEG towards arbitrary high dimensions in realistic experimental settings.
The initial convergence towards the unknown state is slower with around 25 iterations necessary to reach an infidelity of $10\%$.
This is to be expected as MUBs span the qutrit state space more efficiently compared to the Pauli operator measurements. Tab.~\ref{tab:values} shows details of the performance.

\begin{figure}[t]
\includegraphics[width=.95\columnwidth]{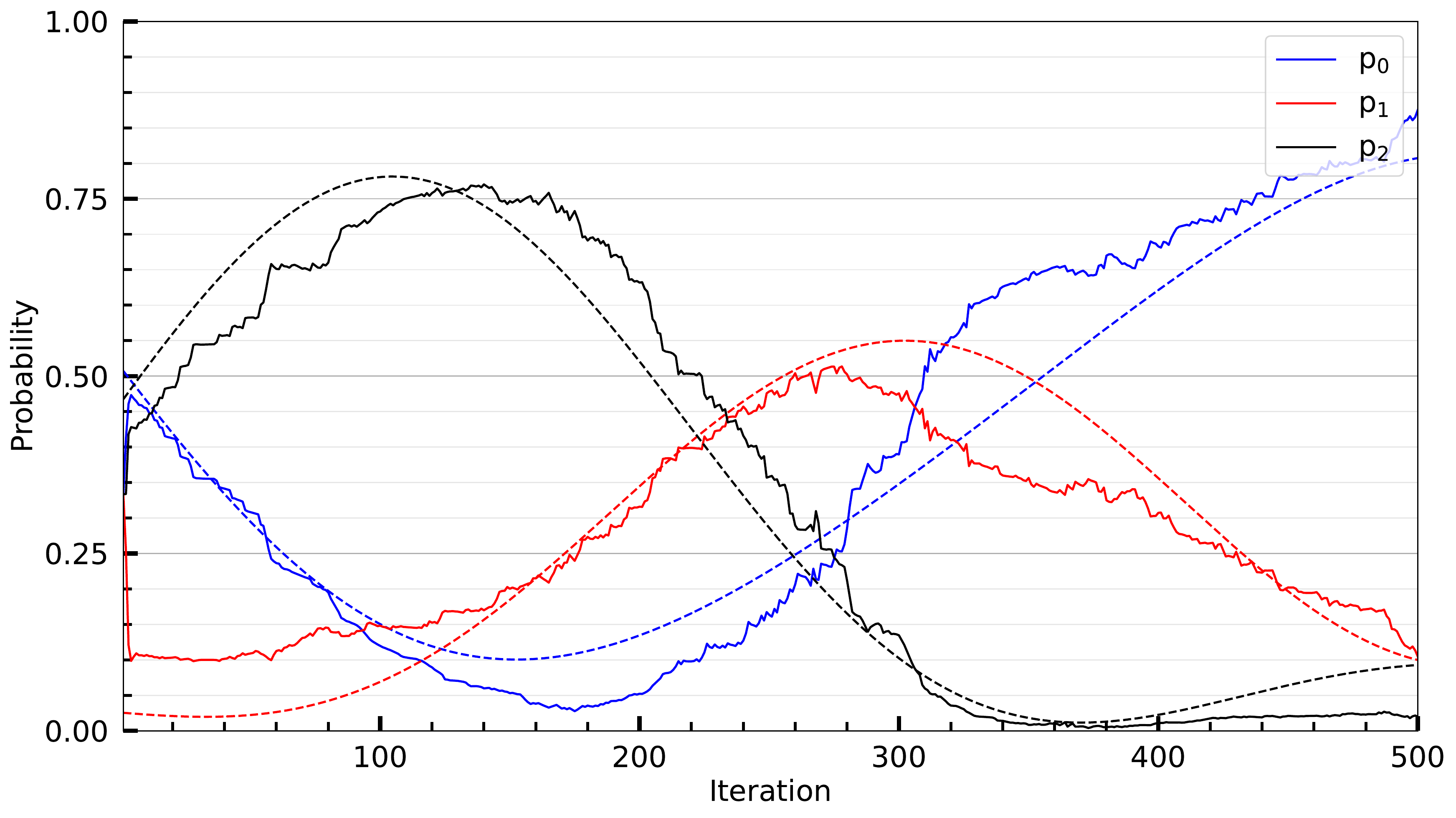}
\caption{
\label{fig:probabilities} 
Tracking of a quantum state under randomly chosen Hermitian evolution $\sigma_r$.
MEG can track the theoretical (dashed) deterministic evolution of a state experimentally (solid).
The three probability components p$_i$ for the logical basis states $i=0,1,2$ (blue, red, black) follow the theory nicely.
Starting at an initial guess of a completely mixed state, useful predictions for p$_i$ are obtained very effectively after 3-4 iterations.
}
\end{figure}

Since a changing state leads to changing probabilities of finding a system in a particular state, we also show that we are able to follow these probabilities. 
Fig.~\ref{fig:probabilities} illustrates one example of these probabilities p$_i = \left| \braket{i|\psi_t} \right|^2$ for the logical basis states ${\ket{i},i=0,1,2}$, out of the 50 randomly chosen qutrits analysed in Fig.~\ref{fig:MEG_all}. 
The theoretically expected probabilities in the experiment (dashed lines) are followed nicely by online state tracking of the deterministic evolution in the experiment (solid lines). 
This confirms that the low infidelities directly correspond to the capability to predict measurement outcomes on the states well. 
We also highlight that MEG quickly converges to a useful estimate (after 3-4 iterations), even though the initial guess is a completely mixed qutrit (p$_i = 1/3\,\forall\,i$) to avoid bias in the results.
The small offset between theory and experiment could be further decreased by using an adaptive learning rate.

\begin{table}[b]
\caption{
\label{tab:values} 
MEG performance indicators. 
Top: iterations needed to reach $10\%$ infidelity. 
Bottom: mean infidelity. 
All values achieved with a constant learning rate $\eta = 5$.
Uncertainties are the boundaries of the interquartile ranges.
}
\begin{tabular}{|c|c|c|c|c|c|c|}
\hline
\multirow{2}{*}{\hspace{1cm}\textbf{Signal rate (Hz)}\hspace{1cm}} & \multicolumn{3}{c|}{\textbf{MUB}} & \multicolumn{3}{c|}{\textbf{Pauli operators}} \\ 
\cline{2-7}
& \hspace{.2cm} $\sigma = 0_{3,3}$ \hspace{.2cm} & \hspace{.2cm} $\sigma = \sigma_z$ \hspace{.2cm} & \hspace{.2cm} $\sigma = \sigma_{r}$ \hspace{.2cm} & \hspace{.2cm} $\sigma = 0_{3,3}$ \hspace{.2cm} & \hspace{.2cm} $\sigma = \sigma_z$ \hspace{.2cm} & \hspace{.2cm} $\sigma = \sigma_{r}$ \hspace{.2cm} \\ 
\hline \hline
& \multicolumn{6}{c|}{\textbf{Iterations for Infidelity $<10\%$}} \\ 
\hline
$10^2$ & $14~\substack{+17 \\ -10}$ & $13~\substack{+26 \\ -6}$ & $16~\substack{+26 \\ -12}$ & $30~\substack{+32 \\ -21}$ & $23~\substack{+68 \\ -14}$ & $24~\substack{+38 \\ -17}$ \\
$10^6$ & $4~\substack{+2 \\ -1}$ &$4~\substack{+3 \\ -1}$ &$4~\substack{+4 \\ -1}$ & && \\
\hline \hline
& \multicolumn{6}{c|}{\textbf{Mean Infidelity}} \\ 
\hline 
$10^2$ & $5.3~\substack{+3.2 \\ -2.2}$ & $5.4~\substack{+3.4 \\ -2.4}$ & $5.6~\substack{+3.4 \\ -2.5}$ & $4.9~\substack{+2.8 \\ -2.0}$ & $5.4~\substack{+3.5 \\ -2.4}$ & $5.0~\substack{+3.3 \\ -2.1}$ \\
$10^6$ & $3.4~\substack{+1.8 \\ -1.2}$ &$5.1~\substack{+2.2 \\ -1.9}$ &$4.7~\substack{+2.3 \\ -1.9}$ & && \\
\hline
\end{tabular}

\end{table}

\subsection{Noisy MEG}

\begin{figure}[t]
\includegraphics[width=.95\columnwidth]{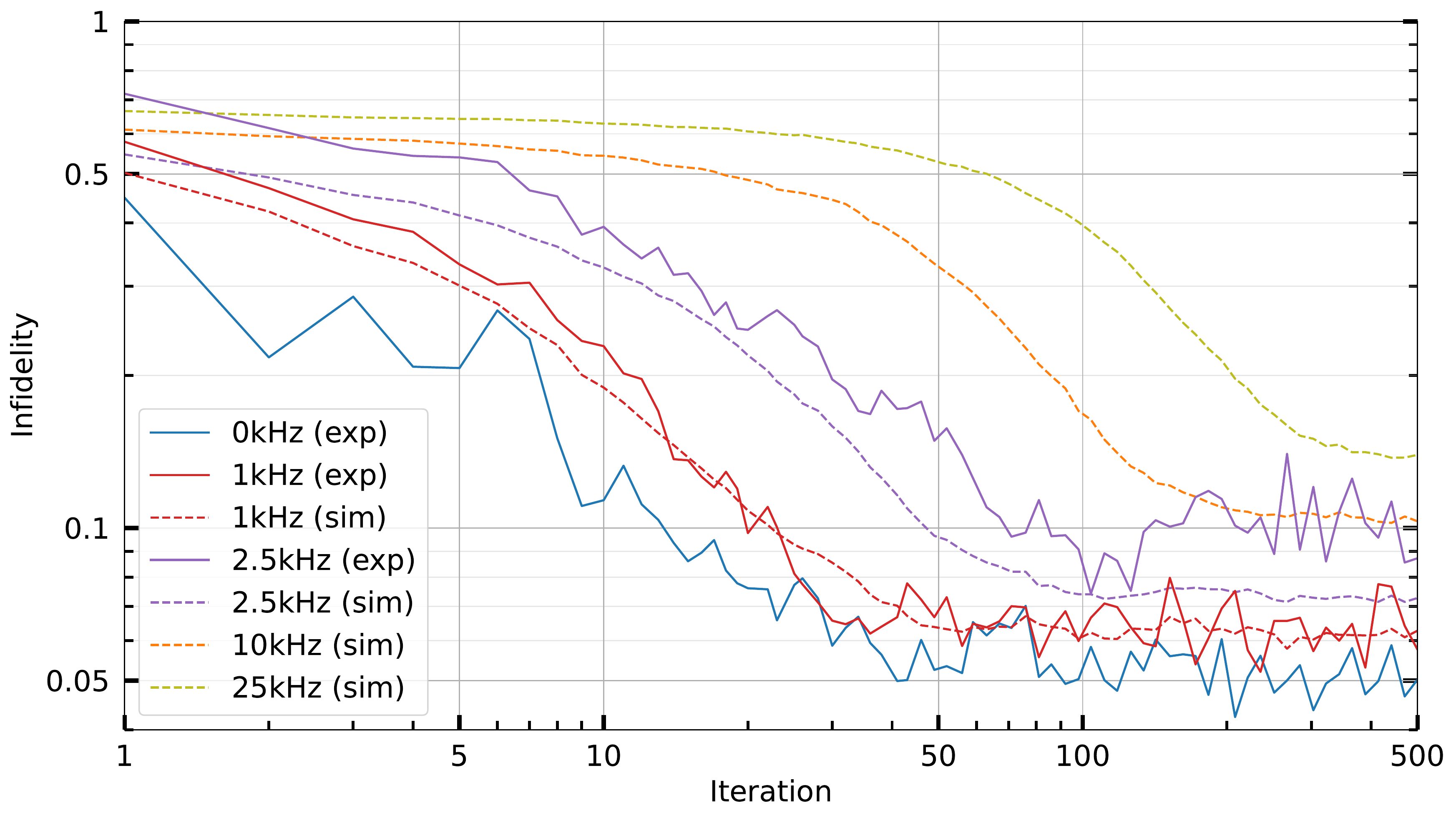}
\caption{
\label{fig:noise_single} 
MEG performance under excessive noise in experiment (exp, solid) and simulation (sim, dashed) with low signal rates ($\sim 10^2$ photons per measurement), using MUBs.
Median performance out of 50 randomly chosen states $\ket{\psi_0}$ (experiment) and 20 runs each (total of $50\times20=1000$ traces) of added Poissonian distributed noise (simulation).
Uncertainty ranges omitted for clarity.
}
\end{figure}

\noindent 
We tested the efficacy of MEG under significant environmental noise experimentally---by adding a light bulb close to the detector while keeping the signal rate at $N \approx 10^2$ photons per measurement---and in simulations---adding Poissonian distributed noise to the raw data. 
The strength of the noise in the experiment is controlled via the distance and angle of the light bulb to the single photon detector.
As a proof of principle demonstration, we only looked at stationary states using MUBs in our experiment, yet the simulations show similar behaviour for all estimation and online tracking scenarios investigated in the previous section.
The median infidelities of 50 randomly chosen states for different methods (solid---experiment, dashed---simulation) and noise strengths (colours) are shown in Fig.~\ref{fig:noise_single}.
Shaded uncertainty ranges similar to Fig.~\ref{fig:MEG_all} are omitted for clarity.
As expected, the achievable infidelity and number of iterations to get there both become higher as the noise increases, however, MEG is showing exceptional robustness.

We observe that up to $N_{back} \approx 2$\,kHz of experimental and simulated noise (not shown explicitly in the figure), the infidelities are comparable to the \textit{no noise} case within uncertainty bars. 
This corresponds to a signal-to-noise ratio SNR\,$\approx$\,0.05, so only one in 20 photons carry the actual information on the unknown quantum state prepared in the experiment.
It is noteworthy that even at 0~kHz of artificially introduced noise, the naturally occurring counting noise together with background and dark count rates are almost double the signal count rate, i.e., $\sim170$ compared to $\sim100$ photons per measurement. 
Experimental and simulated data for $N_{back} = \left[1,2.5\right]$\,kHz show good overlap within uncertainty bars, indicating the validity of the noise model in simulation.
Therefore we expect to achieve infidelities of around $10\%$ up to SNR\,$\approx$\,0.01 (orange dashed line).
The seemingly smoother performance of the simulations stems from 20 runs of randomly added shot-to-shot noise for each of the 50 investigated states.
In other words, the results from simulations reflect the median infidelity of $50\times 20 = 1000$ traces, compared to 50 in the experiment.

\section{Discussion}

\noindent 
We have presented an efficient tomography algorithm to track the continuous evolution of quantum states despite significant environmental noise. 
Our approach, matrix-exponentiated gradient tomography (MEG), uses machine learning techniques to iteratively update an estimate of a quantum state in real time.
We have extended the original online tomography proposal~\cite{Youssry2019} to arbitrary high-dimensional systems (qudits) and experimentally demonstrated the algorithm with a qutrit encoded in the shape of photons.
We have shown that MEG is highly resilient to noise: we are able to estimate the quantum state with high fidelities, and accurately predict probabilities of measurement outcomes for different types of evolution and signal strengths. 
Our technique will enable applications that benefit from efficient online tomography, e.g., quantum process tomography~\cite{Chuang1997a,OBrien2004a} and quantum error mitigation~\cite{Endo2018}/correction~\cite{Lidar2013}.

Our results can be extended to process tomography through the Choi-Jamio\l{}kowski isomorphism~\cite{Jamiolkowski1972,Choi1975}, which allows us to represent a channel using a quantum state that lives in a $d^2$-dimensional Hilbert space. 
In this case, the process estimation becomes state estimation, and MEG could be utilised for the procedure. 
This provides a more efficient solution compared to standard process tomography which requires computational complexity $O(d^8)$ by using a tomographically complete set of states in preparation and measurement, even $O(d^{10})$ for an over-complete set~\cite{Knee2018}. 
The bottleneck of the standard procedure comes from the matrix-vector multiplications required for the estimation procedure. On the other hand, for the MEG update rule, the bottleneck operation is the matrix exponential which is computationally more efficient (see~\cite{Youssry2019} and references within). 

\section{Methods}

\subsection{Experimental Setup and Measurement}

\noindent 
The quantum states are encoded in the transverse spatial mode---the shape---of single photons.
We describe these modes in the Laguerre-Gaussian basis $\{\ket{l_i,p_i}\}$~\cite{Allen2003}, where each randomly chosen state is given by the superposition $\ket{\psi} {=} \sum_i c_i \ket{l_i,p_i}$ (with $\sum_i |c_i|^2 = 1$). 
For the investigated qutrits in the experiment we use the three logical basis states $\{\ket{0}\equiv\ket{-1,0}, \ket{1}\equiv\ket{0,2}, \ket{2}\equiv\ket{1,0}\}$, keeping the Gouy phase of the same order and therefore preventing the shape from rotating as the photons propagate~\cite{Langford2004}.
The experiment is conducted with highly attenuated CW laser light (Thorlabs
, mean photon number $|\alpha|^2 = 0.01$) together with two spatial light modulators (SLM, Meadowlark Optics
) and a single photon detector (Perkin-Elmer
, $\sim 100$\,Hz dark count rate).
The first SLM is used to prepare the \textit{unknown} quantum state $\ket{\psi_t}$ via a hologram that changes phase and amplitude of the initial state $\ket{l=0,p=0}$ from the laser source.
A second hologram displayed on the other SLM is then measuring the state $\ket{\psi_m}$ from one randomly chosen basis set (see \textit{MEG Estimation Extensions} section for details).
Finally, a single-mode fibre and the detector are acting as a local filter with the number of detected photons proportional to the overlap $\left| \braket{\psi_m|\psi_t} \right|^2$ between the prepared and measured mode.
The schematic, details, and characterisation of the simple but effective prepare-and-measure experiment are described extensively in Ref.~\cite{Rambach2021} and its supplemental material.

\subsection{MEG Estimation Extensions}

\subsubsection{Measurement Schemes}\label{sec:Meas}
\noindent 
We use two measurement schemes to perform our state estimation and online tracking. 
The first scheme is utilising Mutually Unbiased Bases (MUBs)~\cite{Durt2010}, measurements that are convenient to perform on the photonic system in our experiments. 
For a $d$-dimensional quantum system, we have $d+1$ basis sets, each set consists of $d$ states. 
However, complete sets of MUBs are not know for every dimension and so, we also introduced a second scheme using generalised Pauli operators~\cite{Kimura2003}. 
These operators are given by the set $\{u_{jk}, v_{jk}, w_{l}\}$
\begin{subequations}
\begin{align}
    u_{jk} &= \ket{j}\! \!\bra{k} + \ket{k}\!\!\bra{j}, && 0\le j\le k\le d \\
    v_{jk} &= -i\ket{j}\!\!\bra{k} +i\ket{k}\!\!\bra{j}, && 0\le j\le k\le d \\
    w_l &= \sqrt{\frac{2}{l+1}}\left(-l\ket{l}\!\!\bra{l} + \sum_{j=0}^{l}{\ket{j}\!\!\bra{j}} \right). && 0\le l \le d
\end{align}
\end{subequations}
The operators reduce to the Pauli and Gell-Mann matrices for $d=2$ and $d=3$, respectively. The total number of operators is $d^2 -1$, and for each operator, there are $d$ eigenstates that could be measured.
The generalised Pauli operators form an orthonormal basis set, and thus are informationally-complete.

\subsubsection{Modified update rule}
\noindent 
The learning rate is usually chosen to be adaptive to guarantee convergence in the case of noisy measurements as shown in \cite{Youssry2019}. 
However, upon exploring the effect of choosing different learning rates, we found that a constant learning rate is sufficient to obtain an estimate with the maximum possible accuracy obtainable given our experimental capabilities. 
Thus, we fix the learning rate to be $\eta_t=\eta=5$. 

The measurement schemes we use provide us with more information at each iteration. 
In the case of MUB measurements, we randomly select a basis at each iteration. 
In order to calculate the probabilities, we need to projectively measure all states of that basis for normalisation of the measurement counts. 
The same holds when we use generalised Pauli measurements, we still need to measure all the eigenstates of a given Pauli operator. 
Therefore, we modify the loss function to include multiple measurement results obtained at a given iteration as follows. 
\begin{align}
    L_t = \sum_{i=1}^{d}{\left(\tr(\hat{\rho}_{t}X_t^{(i)}) - y_t^{(i)}\right)^2},
\end{align}
with gradient 
\begin{align}
    \nabla L_t = 2\sum_{i=1}^{{d}}\left(\tr(\hat{\rho}_{t}X_t^{(i)}) - y_t^{(i)}\right)X_t^{(i)},
\end{align}
where $X_t^{(i)} = \ket{\psi_i}\!\!\bra{\psi_i}$ is the projector of the $i^{\text{th}}$ state of the MUB (or Pauli operator) basis that was randomly chosen at iteration $t$ out of $d+1$ (or $d^2-1$) possible bases, and $y_t^{(i)}$ is the measured probability. 

Since we are restricting the experiments to pure states in our investigations, we can increase the estimation accuracy using this prior information. 
The method we use is to project the state into the subspace of its largest eigenvalue, or equivalently, finding the pure state closest to our estimate in fidelity.
We illustrate this method with an example: we can write the estimated density matrix $\hat{\rho}_t$ of a pure quantum state $\ket{\phi_t}$ in the form
\begin{align}
    \hat{\rho}_t = \lambda\ket{\phi_t}\!\!\bra{\phi_t} + (1-\lambda) \hat{\rho}_{\text{noise}},
\end{align}
where the parameter $\lambda \in [0,1]$ indicates the noise strength, and $\hat{\rho}_{\text{noise}}=I_d/d$ is the completely mixed state representing the noise in the estimate, with $I_d$ the $d$ dimensional identity matrix. 
The special cases $\lambda=0$ and $\lambda=1$ correspond to the noiseless and completely noisy states respectively. Now, let $\ket{\gamma_t}$ be an eigenstate of $\hat{\rho}_t$ with eigenvalue $\gamma_t$, i.e., $\hat{\rho}_t \ket{\gamma_t}=\gamma_t\ket{\gamma_t}$. 
Then, we can write
\begin{subequations}
\begin{align}
    \gamma_t &= \braket{\gamma_t|\hat{\rho}_t|\gamma_t}\\ &= \lambda\braket{\gamma_t|\phi_t}\braket{\phi_t|\gamma_t} + (1-\lambda) \frac{\braket{\gamma_t|I_d|\gamma_t}}{d} \\
    &= \lambda\left|\braket{\gamma_t|\phi_t}\right|^2 + \frac{1-\lambda}{d}.
\end{align}
\end{subequations}
From there, we can see that the maximum eigenvalue is
\begin{align}
    \gamma_{t, \text{max}} = \lambda\ + \frac{1-\lambda}{d},
\end{align}
which is obtainable when there is maximum overlap between $\gamma_t$ and $\phi_t$, in the best case $\left|\braket{\gamma_t|\phi_t}\right|^2=1$, i.e.,  $\ket{\gamma_t}=\ket{\phi_t}$. 
This shows that the true pure state $\ket{\phi_t}$ is an eigenvector of the density matrix $\hat{\rho_t}$ corresponding to the maximum eigenvalue, in other words, closest to our estimator in fidelity. 
So, after each iteration we calculate the eigenvalues of the state estimate and select the corresponding eigenvector to be the estimate of the pure state $\ket{\phi_t}$.

\section{Data Availability}
\noindent 
All the generated datasets used in this study are publicly available on Figshare at \url{https://figshare.com/projects/Efficient_Tracking_of_Noisy_Quantum_States/130433}. 

\section{Code Availability}
\noindent 
The source code for the proposed theoretical methods is publicly available on github at \url{https://github.com/akramyoussry/MEG_state_tracking}. 

\section{Acknowledgements}
\noindent 
This research was supported by the Australian Research Council Centre of Excellence for Engineered Quantum Systems (EQUS, CE170100009) and Discovery Project (DP200102273). 
AY is supported by an Australian Government Research Training Program Scholarship, and the Australian Research Council under the Centre of Excellence Scheme No. CE170100012.
MT is supported by the National Research Foundation, Prime Minister’s Office, Singapore and the Ministry of Education, Singapore under the Research Centres of Excellence programme, and by NUS startup grants (R-263-000-E32-133 and R-263-000-E32-731).
JR is supported by a Westpac Bicentennial Foundation Research Fellowship. 

\section{Author Contributions}
\noindent 
MR performed the optical experiments and acquired and evaluated the data with inputs from JR. 
MR and AY performed the numerical experiments. 
AY and MT designed and AY implemented the algorithm based on feedback from MR and JR. 
MT and JR supervised the project. 
MR led the writing of the article with contributions from all authors.

\section{Competing Interests}
\noindent 
The authors declare no competing financial or non-financial interests.

\bibliography{MEG_bib}

\begin{thebibliography}{36}%
\makeatletter
\providecommand \@ifxundefined [1]{%
 \@ifx{#1\undefined}
}%
\providecommand \@ifnum [1]{%
 \ifnum #1\expandafter \@firstoftwo
 \else \expandafter \@secondoftwo
 \fi
}%
\providecommand \@ifx [1]{%
 \ifx #1\expandafter \@firstoftwo
 \else \expandafter \@secondoftwo
 \fi
}%
\providecommand \natexlab [1]{#1}%
\providecommand \enquote  [1]{``#1''}%
\providecommand \bibnamefont  [1]{#1}%
\providecommand \bibfnamefont [1]{#1}%
\providecommand \citenamefont [1]{#1}%
\providecommand \href@noop [0]{\@secondoftwo}%
\providecommand \href [0]{\begingroup \@sanitize@url \@href}%
\providecommand \@href[1]{\@@startlink{#1}\@@href}%
\providecommand \@@href[1]{\endgroup#1\@@endlink}%
\providecommand \@sanitize@url [0]{\catcode `\\12\catcode `\$12\catcode
  `\&12\catcode `\#12\catcode `\^12\catcode `\_12\catcode `\%12\relax}%
\providecommand \@@startlink[1]{}%
\providecommand \@@endlink[0]{}%
\providecommand \url  [0]{\begingroup\@sanitize@url \@url }%
\providecommand \@url [1]{\endgroup\@href {#1}{\urlprefix }}%
\providecommand \urlprefix  [0]{URL }%
\providecommand \Eprint [0]{\href }%
\providecommand \doibase [0]{https://doi.org/}%
\providecommand \selectlanguage [0]{\@gobble}%
\providecommand \bibinfo  [0]{\@secondoftwo}%
\providecommand \bibfield  [0]{\@secondoftwo}%
\providecommand \translation [1]{[#1]}%
\providecommand \BibitemOpen [0]{}%
\providecommand \bibitemStop [0]{}%
\providecommand \bibitemNoStop [0]{.\EOS\space}%
\providecommand \EOS [0]{\spacefactor3000\relax}%
\providecommand \BibitemShut  [1]{\csname bibitem#1\endcsname}%
\let\auto@bib@innerbib\@empty
\bibitem [{\citenamefont {Paris}\ and\ \citenamefont
  {Řeh{\'{a}}{\v{c}}ek}(2004)}]{Paris2004}%
  \BibitemOpen
  \bibinfo {editor} {\bibfnamefont {M.}~\bibnamefont {Paris}}\ and\ \bibinfo
  {editor} {\bibfnamefont {J.}~\bibnamefont {Řeh{\'{a}}{\v{c}}ek}},\ eds.,\
  \href {https://doi.org/10.1007/b98673} {\emph {\bibinfo {title} {{Quantum
  State Estimation}}}},\ \bibinfo {series} {Lecture Notes in Physics}, Vol.\
  \bibinfo {volume} {649}\ (\bibinfo  {publisher} {Springer Berlin,
  Heidelberg},\ \bibinfo {year} {2004})\BibitemShut {NoStop}%
\bibitem [{\citenamefont {Haah}\ \emph {et~al.}(2017)\citenamefont {Haah},
  \citenamefont {Harrow}, \citenamefont {Ji}, \citenamefont {Wu},\ and\
  \citenamefont {Yu}}]{Haah17}%
  \BibitemOpen
  \bibfield  {author} {\bibinfo {author} {\bibfnamefont {J.}~\bibnamefont
  {Haah}}, \bibinfo {author} {\bibfnamefont {A.~W.}\ \bibnamefont {Harrow}},
  \bibinfo {author} {\bibfnamefont {Z.}~\bibnamefont {Ji}}, \bibinfo {author}
  {\bibfnamefont {X.}~\bibnamefont {Wu}},\ and\ \bibinfo {author}
  {\bibfnamefont {N.}~\bibnamefont {Yu}},\ }\bibfield  {title} {\bibinfo
  {title} {Sample-optimal tomography of quantum states},\ }\href
  {https://doi.org/10.1109/TIT.2017.2719044} {\bibfield  {journal} {\bibinfo
  {journal} {IEEE Transactions on Information Theory}\ }\textbf {\bibinfo
  {volume} {63}},\ \bibinfo {pages} {5628} (\bibinfo {year}
  {2017})}\BibitemShut {NoStop}%
\bibitem [{\citenamefont {{D'Ariano}}\ \emph {et~al.}(2003)\citenamefont
  {{D'Ariano}}, \citenamefont {Paris},\ and\ \citenamefont
  {Sacchi}}]{Dariano2003}%
  \BibitemOpen
  \bibfield  {author} {\bibinfo {author} {\bibfnamefont {G.~M.}\ \bibnamefont
  {{D'Ariano}}}, \bibinfo {author} {\bibfnamefont {M.~G.}\ \bibnamefont
  {Paris}},\ and\ \bibinfo {author} {\bibfnamefont {M.~F.}\ \bibnamefont
  {Sacchi}},\ }\bibfield  {title} {\bibinfo {title} {{Quantum Tomography}},\
  }in\ \href {https://doi.org/10.1016/S1076-5670(03)80065-4} {\emph {\bibinfo
  {booktitle} {Advances in Imaging and Electron Physics}}},\ Vol.\ \bibinfo
  {volume} {128},\ \bibinfo {editor} {edited by\ \bibinfo {editor}
  {\bibfnamefont {P.~W.}\ \bibnamefont {Hawkes}}}\ (\bibinfo  {publisher}
  {Elsevier},\ \bibinfo {year} {2003})\ pp.\ \bibinfo {pages}
  {205--308}\BibitemShut {NoStop}%
\bibitem [{\citenamefont {Opatrn{\'{y}}}\ \emph {et~al.}(1997)\citenamefont
  {Opatrn{\'{y}}}, \citenamefont {Welsch},\ and\ \citenamefont
  {Vogel}}]{Opatrny1997}%
  \BibitemOpen
  \bibfield  {author} {\bibinfo {author} {\bibfnamefont {T.}~\bibnamefont
  {Opatrn{\'{y}}}}, \bibinfo {author} {\bibfnamefont {D.-G.}\ \bibnamefont
  {Welsch}},\ and\ \bibinfo {author} {\bibfnamefont {W.}~\bibnamefont
  {Vogel}},\ }\bibfield  {title} {\bibinfo {title} {{Least-squares inversion
  for density-matrix reconstruction}},\ }\href
  {https://doi.org/10.1103/PhysRevA.56.1788} {\bibfield  {journal} {\bibinfo
  {journal} {Phys. Rev. A}\ }\textbf {\bibinfo {volume} {56}},\ \bibinfo
  {pages} {1788} (\bibinfo {year} {1997})}\BibitemShut {NoStop}%
\bibitem [{\citenamefont {Aaronson}\ \emph {et~al.}(2019)\citenamefont
  {Aaronson}, \citenamefont {Chen}, \citenamefont {Hazan}, \citenamefont
  {Kale},\ and\ \citenamefont {Nayak}}]{Aaronson2019}%
  \BibitemOpen
  \bibfield  {author} {\bibinfo {author} {\bibfnamefont {S.}~\bibnamefont
  {Aaronson}}, \bibinfo {author} {\bibfnamefont {X.}~\bibnamefont {Chen}},
  \bibinfo {author} {\bibfnamefont {E.}~\bibnamefont {Hazan}}, \bibinfo
  {author} {\bibfnamefont {S.}~\bibnamefont {Kale}},\ and\ \bibinfo {author}
  {\bibfnamefont {A.}~\bibnamefont {Nayak}},\ }\bibfield  {title} {\bibinfo
  {title} {{Online learning of quantum states}},\ }\href
  {https://doi.org/10.1088/1742-5468/ab3988} {\bibfield  {journal} {\bibinfo
  {journal} {J. Stat. Mech. Theory Exp.}\ }\textbf {\bibinfo {volume} {2019}},\
  \bibinfo {pages} {124019} (\bibinfo {year} {2019})}\BibitemShut {NoStop}%
\bibitem [{\citenamefont {Granade}\ \emph {et~al.}(2016)\citenamefont
  {Granade}, \citenamefont {Combes},\ and\ \citenamefont {Cory}}]{Granade2016}%
  \BibitemOpen
  \bibfield  {author} {\bibinfo {author} {\bibfnamefont {C.}~\bibnamefont
  {Granade}}, \bibinfo {author} {\bibfnamefont {J.}~\bibnamefont {Combes}},\
  and\ \bibinfo {author} {\bibfnamefont {D.~G.}\ \bibnamefont {Cory}},\
  }\bibfield  {title} {\bibinfo {title} {{Practical Bayesian tomography}},\
  }\href {https://doi.org/10.1088/1367-2630/18/3/033024} {\bibfield  {journal}
  {\bibinfo  {journal} {New J. Phys.}\ }\textbf {\bibinfo {volume} {18}},\
  \bibinfo {pages} {033024} (\bibinfo {year} {2016})}\BibitemShut {NoStop}%
\bibitem [{\citenamefont {Lin}\ \emph {et~al.}(2020)\citenamefont {Lin},
  \citenamefont {Hsu},\ and\ \citenamefont {Li}}]{Lin2020}%
  \BibitemOpen
  \bibfield  {author} {\bibinfo {author} {\bibfnamefont {C.-M.}\ \bibnamefont
  {Lin}}, \bibinfo {author} {\bibfnamefont {Y.-M.}\ \bibnamefont {Hsu}},\ and\
  \bibinfo {author} {\bibfnamefont {Y.-H.}\ \bibnamefont {Li}},\ }\href
  {http://arxiv.org/abs/2012.15498} {\bibinfo {title} {{An Online Algorithm for
  Maximum-Likelihood Quantum State Tomography}}} (\bibinfo {year} {2020}),\
  \Eprint {https://arxiv.org/abs/2012.15498} {arXiv:2012.15498} \BibitemShut
  {NoStop}%
\bibitem [{\citenamefont {Qi}\ \emph {et~al.}(2017)\citenamefont {Qi},
  \citenamefont {Hou}, \citenamefont {Wang}, \citenamefont {Dong},
  \citenamefont {Zhong}, \citenamefont {Li}, \citenamefont {Xiang},
  \citenamefont {Wiseman}, \citenamefont {Li},\ and\ \citenamefont
  {Guo}}]{Qi2017a}%
  \BibitemOpen
  \bibfield  {author} {\bibinfo {author} {\bibfnamefont {B.}~\bibnamefont
  {Qi}}, \bibinfo {author} {\bibfnamefont {Z.}~\bibnamefont {Hou}}, \bibinfo
  {author} {\bibfnamefont {Y.}~\bibnamefont {Wang}}, \bibinfo {author}
  {\bibfnamefont {D.}~\bibnamefont {Dong}}, \bibinfo {author} {\bibfnamefont
  {H.-S.}\ \bibnamefont {Zhong}}, \bibinfo {author} {\bibfnamefont
  {L.}~\bibnamefont {Li}}, \bibinfo {author} {\bibfnamefont {G.-Y.}\
  \bibnamefont {Xiang}}, \bibinfo {author} {\bibfnamefont {H.~M.}\ \bibnamefont
  {Wiseman}}, \bibinfo {author} {\bibfnamefont {C.-F.}\ \bibnamefont {Li}},\
  and\ \bibinfo {author} {\bibfnamefont {G.-C.}\ \bibnamefont {Guo}},\
  }\bibfield  {title} {\bibinfo {title} {{Adaptive quantum state tomography via
  linear regression estimation: Theory and two-qubit experiment}},\ }\href
  {https://doi.org/10.1038/s41534-017-0016-4} {\bibfield  {journal} {\bibinfo
  {journal} {npj Quantum Inf.}\ }\textbf {\bibinfo {volume} {3}},\ \bibinfo
  {pages} {19} (\bibinfo {year} {2017})}\BibitemShut {NoStop}%
\bibitem [{\citenamefont {Chen}\ and\ \citenamefont {Wang}(2020)}]{Chen2020}%
  \BibitemOpen
  \bibfield  {author} {\bibinfo {author} {\bibfnamefont {Y.}~\bibnamefont
  {Chen}}\ and\ \bibinfo {author} {\bibfnamefont {X.}~\bibnamefont {Wang}},\
  }\href {http://arxiv.org/abs/2006.01013} {\bibinfo {title} {{More Practical
  and Adaptive Algorithms for Online Quantum State Learning}}} (\bibinfo {year}
  {2020}),\ \Eprint {https://arxiv.org/abs/2006.01013} {arXiv:2006.01013}
  \BibitemShut {NoStop}%
\bibitem [{\citenamefont {Nohara}\ \emph {et~al.}(2020)\citenamefont {Nohara},
  \citenamefont {Okamoto}, \citenamefont {Fujiwara},\ and\ \citenamefont
  {Takeuchi}}]{Nohara2020}%
  \BibitemOpen
  \bibfield  {author} {\bibinfo {author} {\bibfnamefont {S.}~\bibnamefont
  {Nohara}}, \bibinfo {author} {\bibfnamefont {R.}~\bibnamefont {Okamoto}},
  \bibinfo {author} {\bibfnamefont {A.}~\bibnamefont {Fujiwara}},\ and\
  \bibinfo {author} {\bibfnamefont {S.}~\bibnamefont {Takeuchi}},\ }\bibfield
  {title} {\bibinfo {title} {{Adaptive quantum state estimation for dynamic
  quantum states}},\ }\href {https://doi.org/10.1103/PhysRevA.102.030401}
  {\bibfield  {journal} {\bibinfo  {journal} {Phys. Rev. A}\ }\textbf {\bibinfo
  {volume} {102}},\ \bibinfo {pages} {030401} (\bibinfo {year}
  {2020})}\BibitemShut {NoStop}%
\bibitem [{\citenamefont {Quek}\ \emph {et~al.}(2021)\citenamefont {Quek},
  \citenamefont {Fort},\ and\ \citenamefont {Ng}}]{Quek2021a}%
  \BibitemOpen
  \bibfield  {author} {\bibinfo {author} {\bibfnamefont {Y.}~\bibnamefont
  {Quek}}, \bibinfo {author} {\bibfnamefont {S.}~\bibnamefont {Fort}},\ and\
  \bibinfo {author} {\bibfnamefont {H.~K.}\ \bibnamefont {Ng}},\ }\bibfield
  {title} {\bibinfo {title} {{Adaptive quantum state tomography with neural
  networks}},\ }\href {https://doi.org/10.1038/s41534-021-00436-9} {\bibfield
  {journal} {\bibinfo  {journal} {npj Quantum Inf.}\ }\textbf {\bibinfo
  {volume} {7}},\ \bibinfo {pages} {105} (\bibinfo {year} {2021})}\BibitemShut
  {NoStop}%
\bibitem [{\citenamefont {Ferrie}(2014)}]{Ferrie2014}%
  \BibitemOpen
  \bibfield  {author} {\bibinfo {author} {\bibfnamefont {C.}~\bibnamefont
  {Ferrie}},\ }\bibfield  {title} {\bibinfo {title} {{Self-Guided Quantum
  Tomography}},\ }\href {https://doi.org/10.1103/PhysRevLett.113.190404}
  {\bibfield  {journal} {\bibinfo  {journal} {Phys. Rev. Lett.}\ }\textbf
  {\bibinfo {volume} {113}},\ \bibinfo {pages} {190404} (\bibinfo {year}
  {2014})}\BibitemShut {NoStop}%
\bibitem [{\citenamefont {Li}\ and\ \citenamefont {Cevher}(2019)}]{Li2019}%
  \BibitemOpen
  \bibfield  {author} {\bibinfo {author} {\bibfnamefont {Y.-H.}\ \bibnamefont
  {Li}}\ and\ \bibinfo {author} {\bibfnamefont {V.}~\bibnamefont {Cevher}},\
  }\bibfield  {title} {\bibinfo {title} {{Convergence of the Exponentiated
  Gradient Method with Armijo Line Search}},\ }\href
  {https://doi.org/10.1007/s10957-018-1428-9} {\bibfield  {journal} {\bibinfo
  {journal} {J. Optim. Theory Appl.}\ }\textbf {\bibinfo {volume} {181}},\
  \bibinfo {pages} {588} (\bibinfo {year} {2019})}\BibitemShut {NoStop}%
\bibitem [{\citenamefont {Youssry}\ \emph {et~al.}(2019)\citenamefont
  {Youssry}, \citenamefont {Ferrie},\ and\ \citenamefont
  {Tomamichel}}]{Youssry2019}%
  \BibitemOpen
  \bibfield  {author} {\bibinfo {author} {\bibfnamefont {A.}~\bibnamefont
  {Youssry}}, \bibinfo {author} {\bibfnamefont {C.}~\bibnamefont {Ferrie}},\
  and\ \bibinfo {author} {\bibfnamefont {M.}~\bibnamefont {Tomamichel}},\
  }\bibfield  {title} {\bibinfo {title} {{Efficient online quantum state
  estimation using a matrix-exponentiated gradient method}},\ }\href
  {https://doi.org/10.1088/1367-2630/ab0438} {\bibfield  {journal} {\bibinfo
  {journal} {New J. Phys.}\ }\textbf {\bibinfo {volume} {21}},\ \bibinfo
  {pages} {033006} (\bibinfo {year} {2019})}\BibitemShut {NoStop}%
\bibitem [{\citenamefont {Zhang}\ \emph {et~al.}(2020)\citenamefont {Zhang},
  \citenamefont {Cong}, \citenamefont {Li},\ and\ \citenamefont
  {Wang}}]{Zhang2020}%
  \BibitemOpen
  \bibfield  {author} {\bibinfo {author} {\bibfnamefont {K.}~\bibnamefont
  {Zhang}}, \bibinfo {author} {\bibfnamefont {S.}~\bibnamefont {Cong}},
  \bibinfo {author} {\bibfnamefont {K.}~\bibnamefont {Li}},\ and\ \bibinfo
  {author} {\bibfnamefont {T.}~\bibnamefont {Wang}},\ }\bibfield  {title}
  {\bibinfo {title} {{An online optimization algorithm for the real-time
  quantum state tomography}},\ }\href
  {https://doi.org/10.1007/s11128-020-02866-4} {\bibfield  {journal} {\bibinfo
  {journal} {Quantum Inf. Process.}\ }\textbf {\bibinfo {volume} {19}},\
  \bibinfo {pages} {361} (\bibinfo {year} {2020})}\BibitemShut {NoStop}%
\bibitem [{\citenamefont {Silberfarb}\ \emph {et~al.}(2005)\citenamefont
  {Silberfarb}, \citenamefont {Jessen},\ and\ \citenamefont
  {Deutsch}}]{Silberfarb2005}%
  \BibitemOpen
  \bibfield  {author} {\bibinfo {author} {\bibfnamefont {A.}~\bibnamefont
  {Silberfarb}}, \bibinfo {author} {\bibfnamefont {P.~S.}\ \bibnamefont
  {Jessen}},\ and\ \bibinfo {author} {\bibfnamefont {I.~H.}\ \bibnamefont
  {Deutsch}},\ }\bibfield  {title} {\bibinfo {title} {{Quantum State
  Reconstruction via Continuous Measurement}},\ }\href
  {https://doi.org/10.1103/PhysRevLett.95.030402} {\bibfield  {journal}
  {\bibinfo  {journal} {Phys. Rev. Lett.}\ }\textbf {\bibinfo {volume} {95}},\
  \bibinfo {pages} {030402} (\bibinfo {year} {2005})}\BibitemShut {NoStop}%
\bibitem [{\citenamefont {Shabani}\ \emph {et~al.}(2014)\citenamefont
  {Shabani}, \citenamefont {Roden},\ and\ \citenamefont
  {Whaley}}]{Shabani2014}%
  \BibitemOpen
  \bibfield  {author} {\bibinfo {author} {\bibfnamefont {A.}~\bibnamefont
  {Shabani}}, \bibinfo {author} {\bibfnamefont {J.}~\bibnamefont {Roden}},\
  and\ \bibinfo {author} {\bibfnamefont {K.~B.}\ \bibnamefont {Whaley}},\
  }\bibfield  {title} {\bibinfo {title} {{Continuous Measurement of a
  Non-Markovian Open Quantum System}},\ }\href
  {https://doi.org/10.1103/PhysRevLett.112.113601} {\bibfield  {journal}
  {\bibinfo  {journal} {Phys. Rev. Lett.}\ }\textbf {\bibinfo {volume} {112}},\
  \bibinfo {pages} {113601} (\bibinfo {year} {2014})}\BibitemShut {NoStop}%
\bibitem [{\citenamefont {Cole}\ \emph {et~al.}(2005)\citenamefont {Cole},
  \citenamefont {Schirmer}, \citenamefont {Greentree}, \citenamefont {Wellard},
  \citenamefont {Oi},\ and\ \citenamefont {Hollenberg}}]{Cole2005}%
  \BibitemOpen
  \bibfield  {author} {\bibinfo {author} {\bibfnamefont {J.~H.}\ \bibnamefont
  {Cole}}, \bibinfo {author} {\bibfnamefont {S.~G.}\ \bibnamefont {Schirmer}},
  \bibinfo {author} {\bibfnamefont {A.~D.}\ \bibnamefont {Greentree}}, \bibinfo
  {author} {\bibfnamefont {C.~J.}\ \bibnamefont {Wellard}}, \bibinfo {author}
  {\bibfnamefont {D.~K.~L.}\ \bibnamefont {Oi}},\ and\ \bibinfo {author}
  {\bibfnamefont {L.~C.~L.}\ \bibnamefont {Hollenberg}},\ }\bibfield  {title}
  {\bibinfo {title} {{Identifying an experimental two-state Hamiltonian to
  arbitrary accuracy}},\ }\href {https://doi.org/10.1103/PhysRevA.71.062312}
  {\bibfield  {journal} {\bibinfo  {journal} {Phys. Rev. A}\ }\textbf {\bibinfo
  {volume} {71}},\ \bibinfo {pages} {062312} (\bibinfo {year}
  {2005})}\BibitemShut {NoStop}%
\bibitem [{\citenamefont {Anshu}\ \emph {et~al.}(2021)\citenamefont {Anshu},
  \citenamefont {Arunachalam}, \citenamefont {Kuwahara},\ and\ \citenamefont
  {Soleimanifar}}]{Anshu2021}%
  \BibitemOpen
  \bibfield  {author} {\bibinfo {author} {\bibfnamefont {A.}~\bibnamefont
  {Anshu}}, \bibinfo {author} {\bibfnamefont {S.}~\bibnamefont {Arunachalam}},
  \bibinfo {author} {\bibfnamefont {T.}~\bibnamefont {Kuwahara}},\ and\
  \bibinfo {author} {\bibfnamefont {M.}~\bibnamefont {Soleimanifar}},\
  }\bibfield  {title} {\bibinfo {title} {{Sample-efficient learning of
  interacting quantum systems}},\ }\href
  {https://doi.org/10.1038/s41567-021-01232-0} {\bibfield  {journal} {\bibinfo
  {journal} {Nat. Phys.}\ }\textbf {\bibinfo {volume} {17}},\ \bibinfo {pages}
  {931} (\bibinfo {year} {2021})}\BibitemShut {NoStop}%
\bibitem [{\citenamefont {Bogdanov}\ \emph {et~al.}(2016)\citenamefont
  {Bogdanov}, \citenamefont {Bantysh}, \citenamefont {Bogdanova}, \citenamefont
  {Kvasnyy},\ and\ \citenamefont {Lukichev}}]{Bogdanov2016a}%
  \BibitemOpen
  \bibfield  {author} {\bibinfo {author} {\bibfnamefont {Y.~I.}\ \bibnamefont
  {Bogdanov}}, \bibinfo {author} {\bibfnamefont {B.~I.}\ \bibnamefont
  {Bantysh}}, \bibinfo {author} {\bibfnamefont {N.~A.}\ \bibnamefont
  {Bogdanova}}, \bibinfo {author} {\bibfnamefont {A.~B.}\ \bibnamefont
  {Kvasnyy}},\ and\ \bibinfo {author} {\bibfnamefont {V.~F.}\ \bibnamefont
  {Lukichev}},\ }\bibfield  {title} {\bibinfo {title} {{Quantum states
  tomography with noisy measurement channels}},\ }in\ \href
  {https://doi.org/10.1117/12.2267029} {\emph {\bibinfo {booktitle} {Int. Conf.
  Micro- Nano-Electronics 2016}}},\ Vol.\ \bibinfo {volume} {10224},\ \bibinfo
  {editor} {edited by\ \bibinfo {editor} {\bibfnamefont {V.~F.}\ \bibnamefont
  {Lukichev}}\ and\ \bibinfo {editor} {\bibfnamefont {K.~V.}\ \bibnamefont
  {Rudenko}}}\ (\bibinfo {year} {2016})\ p.\ \bibinfo {pages}
  {102242O}\BibitemShut {NoStop}%
\bibitem [{\citenamefont {Ivanova-Rohling}\ \emph {et~al.}(2022)\citenamefont
  {Ivanova-Rohling}, \citenamefont {Rohling},\ and\ \citenamefont
  {Burkard}}]{Ivanova-Rohling2022}%
  \BibitemOpen
  \bibfield  {author} {\bibinfo {author} {\bibfnamefont {V.~N.}\ \bibnamefont
  {Ivanova-Rohling}}, \bibinfo {author} {\bibfnamefont {N.}~\bibnamefont
  {Rohling}},\ and\ \bibinfo {author} {\bibfnamefont {G.}~\bibnamefont
  {Burkard}},\ }\href {https://doi.org/10.48550/arXiv.2203.05677} {\bibinfo
  {title} {{Optimal Quantum State Tomography with Noisy Gates}}} (\bibinfo
  {year} {2022}),\ \Eprint {https://arxiv.org/abs/2203.05677}
  {arXiv:2203.05677} \BibitemShut {NoStop}%
\bibitem [{\citenamefont {Farooq}\ \emph {et~al.}(2022)\citenamefont {Farooq},
  \citenamefont {Khalid}, \citenamefont {ur~Rehman},\ and\ \citenamefont
  {Shin}}]{Farooq2022}%
  \BibitemOpen
  \bibfield  {author} {\bibinfo {author} {\bibfnamefont {A.}~\bibnamefont
  {Farooq}}, \bibinfo {author} {\bibfnamefont {U.}~\bibnamefont {Khalid}},
  \bibinfo {author} {\bibfnamefont {J.}~\bibnamefont {ur~Rehman}},\ and\
  \bibinfo {author} {\bibfnamefont {H.}~\bibnamefont {Shin}},\ }\bibfield
  {title} {\bibinfo {title} {{Robust Quantum State Tomography Method for
  Quantum Sensing}},\ }\href {https://doi.org/10.3390/s22072669} {\bibfield
  {journal} {\bibinfo  {journal} {Sensors}\ }\textbf {\bibinfo {volume} {22}},\
  \bibinfo {pages} {2669} (\bibinfo {year} {2022})}\BibitemShut {NoStop}%
\bibitem [{\citenamefont {Blume-Kohout}(2010)}]{Blume-Kohout2010}%
  \BibitemOpen
  \bibfield  {author} {\bibinfo {author} {\bibfnamefont {R.}~\bibnamefont
  {Blume-Kohout}},\ }\bibfield  {title} {\bibinfo {title} {{Optimal, reliable
  estimation of quantum states}},\ }\href
  {https://doi.org/10.1088/1367-2630/12/4/043034} {\bibfield  {journal}
  {\bibinfo  {journal} {New J. Phys.}\ }\textbf {\bibinfo {volume} {12}},\
  \bibinfo {pages} {043034} (\bibinfo {year} {2010})}\BibitemShut {NoStop}%
\bibitem [{\citenamefont {Durt}\ \emph {et~al.}(2010)\citenamefont {Durt},
  \citenamefont {Englert}, \citenamefont {Bengtsson},\ and\ \citenamefont
  {{\.{Z}}yczkowski}}]{Durt2010}%
  \BibitemOpen
  \bibfield  {author} {\bibinfo {author} {\bibfnamefont {T.}~\bibnamefont
  {Durt}}, \bibinfo {author} {\bibfnamefont {B.-G.}\ \bibnamefont {Englert}},
  \bibinfo {author} {\bibfnamefont {I.}~\bibnamefont {Bengtsson}},\ and\
  \bibinfo {author} {\bibfnamefont {K.}~\bibnamefont {{\.{Z}}yczkowski}},\
  }\bibfield  {title} {\bibinfo {title} {{On Mutually Unbiased Bases}},\ }\href
  {https://doi.org/10.1142/S0219749910006502} {\bibfield  {journal} {\bibinfo
  {journal} {Int. J. Quantum Inf.}\ }\textbf {\bibinfo {volume} {08}},\
  \bibinfo {pages} {535} (\bibinfo {year} {2010})}\BibitemShut {NoStop}%
\bibitem [{\citenamefont {Kimura}(2003)}]{Kimura2003}%
  \BibitemOpen
  \bibfield  {author} {\bibinfo {author} {\bibfnamefont {G.}~\bibnamefont
  {Kimura}},\ }\bibfield  {title} {\bibinfo {title} {{The Bloch vector for
  N-level systems}},\ }\href {https://doi.org/10.1016/S0375-9601(03)00941-1}
  {\bibfield  {journal} {\bibinfo  {journal} {Phys. Lett. A}\ }\textbf
  {\bibinfo {volume} {314}},\ \bibinfo {pages} {339} (\bibinfo {year}
  {2003})}\BibitemShut {NoStop}%
\bibitem [{\citenamefont {Bogdanov}(2009)}]{Bogdanov2009}%
  \BibitemOpen
  \bibfield  {author} {\bibinfo {author} {\bibfnamefont {Y.~I.}\ \bibnamefont
  {Bogdanov}},\ }\bibfield  {title} {\bibinfo {title} {{Unified statistical
  method for reconstructing quantum states by purification}},\ }\href
  {https://doi.org/10.1134/S106377610906003X} {\bibfield  {journal} {\bibinfo
  {journal} {J. Exp. Theor. Phys.}\ }\textbf {\bibinfo {volume} {108}},\
  \bibinfo {pages} {928} (\bibinfo {year} {2009})}\BibitemShut {NoStop}%
\bibitem [{\citenamefont {Chuang}\ and\ \citenamefont
  {Nielsen}(1997)}]{Chuang1997a}%
  \BibitemOpen
  \bibfield  {author} {\bibinfo {author} {\bibfnamefont {I.~L.}\ \bibnamefont
  {Chuang}}\ and\ \bibinfo {author} {\bibfnamefont {M.~A.}\ \bibnamefont
  {Nielsen}},\ }\bibfield  {title} {\bibinfo {title} {{Prescription for
  experimental determination of the dynamics of a quantum black box}},\ }\href
  {https://doi.org/10.1080/09500349708231894} {\bibfield  {journal} {\bibinfo
  {journal} {J. Mod. Opt.}\ }\textbf {\bibinfo {volume} {44}},\ \bibinfo
  {pages} {2455} (\bibinfo {year} {1997})}\BibitemShut {NoStop}%
\bibitem [{\citenamefont {O'Brien}\ \emph {et~al.}(2004)\citenamefont
  {O'Brien}, \citenamefont {Pryde}, \citenamefont {Gilchrist}, \citenamefont
  {James}, \citenamefont {Langford}, \citenamefont {Ralph},\ and\ \citenamefont
  {White}}]{OBrien2004a}%
  \BibitemOpen
  \bibfield  {author} {\bibinfo {author} {\bibfnamefont {J.~L.}\ \bibnamefont
  {O'Brien}}, \bibinfo {author} {\bibfnamefont {G.~J.}\ \bibnamefont {Pryde}},
  \bibinfo {author} {\bibfnamefont {A.}~\bibnamefont {Gilchrist}}, \bibinfo
  {author} {\bibfnamefont {D.~F.~V.}\ \bibnamefont {James}}, \bibinfo {author}
  {\bibfnamefont {N.~K.}\ \bibnamefont {Langford}}, \bibinfo {author}
  {\bibfnamefont {T.~C.}\ \bibnamefont {Ralph}},\ and\ \bibinfo {author}
  {\bibfnamefont {A.~G.}\ \bibnamefont {White}},\ }\bibfield  {title} {\bibinfo
  {title} {{Quantum Process Tomography of a Controlled-NOT Gate}},\ }\href
  {https://doi.org/10.1103/PhysRevLett.93.080502} {\bibfield  {journal}
  {\bibinfo  {journal} {Phys. Rev. Lett.}\ }\textbf {\bibinfo {volume} {93}},\
  \bibinfo {pages} {080502} (\bibinfo {year} {2004})}\BibitemShut {NoStop}%
\bibitem [{\citenamefont {Endo}\ \emph {et~al.}(2018)\citenamefont {Endo},
  \citenamefont {Benjamin},\ and\ \citenamefont {Li}}]{Endo2018}%
  \BibitemOpen
  \bibfield  {author} {\bibinfo {author} {\bibfnamefont {S.}~\bibnamefont
  {Endo}}, \bibinfo {author} {\bibfnamefont {S.~C.}\ \bibnamefont {Benjamin}},\
  and\ \bibinfo {author} {\bibfnamefont {Y.}~\bibnamefont {Li}},\ }\bibfield
  {title} {\bibinfo {title} {{Practical Quantum Error Mitigation for
  Near-Future Applications}},\ }\href
  {https://doi.org/10.1103/PhysRevX.8.031027} {\bibfield  {journal} {\bibinfo
  {journal} {Phys. Rev. X}\ }\textbf {\bibinfo {volume} {8}},\ \bibinfo {pages}
  {031027} (\bibinfo {year} {2018})}\BibitemShut {NoStop}%
\bibitem [{\citenamefont {Lidar}\ and\ \citenamefont {Brun}(2012)}]{Lidar2013}%
  \BibitemOpen
  \bibinfo {editor} {\bibfnamefont {D.~A.}\ \bibnamefont {Lidar}}\ and\
  \bibinfo {editor} {\bibfnamefont {T.~A.}\ \bibnamefont {Brun}},\ eds.,\ \href
  {https://doi.org/10.1017/CBO9781139034807} {\emph {\bibinfo {title} {Quantum
  Error Correction}}}\ (\bibinfo  {publisher} {Cambridge University Press},\
  \bibinfo {address} {Cambridge},\ \bibinfo {year} {2012})\BibitemShut
  {NoStop}%
\bibitem [{\citenamefont {Jamio{\l}kowski}(1972)}]{Jamiolkowski1972}%
  \BibitemOpen
  \bibfield  {author} {\bibinfo {author} {\bibfnamefont {A.}~\bibnamefont
  {Jamio{\l}kowski}},\ }\bibfield  {title} {\bibinfo {title} {{Linear
  transformations which preserve trace and positive semidefiniteness of
  operators}},\ }\href {https://doi.org/10.1016/0034-4877(72)90011-0}
  {\bibfield  {journal} {\bibinfo  {journal} {Reports Math. Phys.}\ }\textbf
  {\bibinfo {volume} {3}},\ \bibinfo {pages} {275} (\bibinfo {year}
  {1972})}\BibitemShut {NoStop}%
\bibitem [{\citenamefont {Choi}(1975)}]{Choi1975}%
  \BibitemOpen
  \bibfield  {author} {\bibinfo {author} {\bibfnamefont {M.-D.}\ \bibnamefont
  {Choi}},\ }\bibfield  {title} {\bibinfo {title} {{Completely positive linear
  maps on complex matrices}},\ }\href
  {https://doi.org/10.1016/0024-3795(75)90075-0} {\bibfield  {journal}
  {\bibinfo  {journal} {Linear Algebra Appl.}\ }\textbf {\bibinfo {volume}
  {10}},\ \bibinfo {pages} {285} (\bibinfo {year} {1975})}\BibitemShut
  {NoStop}%
\bibitem [{\citenamefont {Knee}\ \emph {et~al.}(2018)\citenamefont {Knee},
  \citenamefont {Bolduc}, \citenamefont {Leach},\ and\ \citenamefont
  {Gauger}}]{Knee2018}%
  \BibitemOpen
  \bibfield  {author} {\bibinfo {author} {\bibfnamefont {G.~C.}\ \bibnamefont
  {Knee}}, \bibinfo {author} {\bibfnamefont {E.}~\bibnamefont {Bolduc}},
  \bibinfo {author} {\bibfnamefont {J.}~\bibnamefont {Leach}},\ and\ \bibinfo
  {author} {\bibfnamefont {E.~M.}\ \bibnamefont {Gauger}},\ }\bibfield  {title}
  {\bibinfo {title} {{Quantum process tomography via completely positive and
  trace-preserving projection}},\ }\href
  {https://doi.org/10.1103/PhysRevA.98.062336} {\bibfield  {journal} {\bibinfo
  {journal} {Phys. Rev. A}\ }\textbf {\bibinfo {volume} {98}},\ \bibinfo
  {pages} {062336} (\bibinfo {year} {2018})}\BibitemShut {NoStop}%
\bibitem [{\citenamefont {Allen}\ \emph {et~al.}(2016)\citenamefont {Allen},
  \citenamefont {Barnett},\ and\ \citenamefont {Padgett}}]{Allen2003}%
  \BibitemOpen
  \bibfield  {author} {\bibinfo {author} {\bibfnamefont {L.}~\bibnamefont
  {Allen}}, \bibinfo {author} {\bibfnamefont {S.~M.}\ \bibnamefont {Barnett}},\
  and\ \bibinfo {author} {\bibfnamefont {M.~J.}\ \bibnamefont {Padgett}},\
  }\href {https://doi.org/10.1201/9781482269017} {\emph {\bibinfo {title} {Opt.
  Angular Momentum}}}\ (\bibinfo  {publisher} {CRC Press},\ \bibinfo {address}
  {Boca Raton},\ \bibinfo {year} {2016})\BibitemShut {NoStop}%
\bibitem [{\citenamefont {Langford}\ \emph {et~al.}(2004)\citenamefont
  {Langford}, \citenamefont {Dalton}, \citenamefont {Harvey}, \citenamefont
  {O'Brien}, \citenamefont {Pryde}, \citenamefont {Gilchrist}, \citenamefont
  {Bartlett},\ and\ \citenamefont {White}}]{Langford2004}%
  \BibitemOpen
  \bibfield  {author} {\bibinfo {author} {\bibfnamefont {N.~K.}\ \bibnamefont
  {Langford}}, \bibinfo {author} {\bibfnamefont {R.~B.}\ \bibnamefont
  {Dalton}}, \bibinfo {author} {\bibfnamefont {M.~D.}\ \bibnamefont {Harvey}},
  \bibinfo {author} {\bibfnamefont {J.~L.}\ \bibnamefont {O'Brien}}, \bibinfo
  {author} {\bibfnamefont {G.~J.}\ \bibnamefont {Pryde}}, \bibinfo {author}
  {\bibfnamefont {A.}~\bibnamefont {Gilchrist}}, \bibinfo {author}
  {\bibfnamefont {S.~D.}\ \bibnamefont {Bartlett}},\ and\ \bibinfo {author}
  {\bibfnamefont {A.~G.}\ \bibnamefont {White}},\ }\bibfield  {title} {\bibinfo
  {title} {{Measuring Entangled Qutrits and Their Use for Quantum Bit
  Commitment}},\ }\href {https://doi.org/10.1103/PhysRevLett.93.053601}
  {\bibfield  {journal} {\bibinfo  {journal} {Phys. Rev. Lett.}\ }\textbf
  {\bibinfo {volume} {93}},\ \bibinfo {pages} {053601} (\bibinfo {year}
  {2004})}\BibitemShut {NoStop}%
\bibitem [{\citenamefont {Rambach}\ \emph {et~al.}(2021)\citenamefont
  {Rambach}, \citenamefont {Qaryan}, \citenamefont {Kewming}, \citenamefont
  {Ferrie}, \citenamefont {White},\ and\ \citenamefont {Romero}}]{Rambach2021}%
  \BibitemOpen
  \bibfield  {author} {\bibinfo {author} {\bibfnamefont {M.}~\bibnamefont
  {Rambach}}, \bibinfo {author} {\bibfnamefont {M.}~\bibnamefont {Qaryan}},
  \bibinfo {author} {\bibfnamefont {M.}~\bibnamefont {Kewming}}, \bibinfo
  {author} {\bibfnamefont {C.}~\bibnamefont {Ferrie}}, \bibinfo {author}
  {\bibfnamefont {A.~G.}\ \bibnamefont {White}},\ and\ \bibinfo {author}
  {\bibfnamefont {J.}~\bibnamefont {Romero}},\ }\bibfield  {title} {\bibinfo
  {title} {{Robust and Efficient High-Dimensional Quantum State Tomography}},\
  }\href {https://doi.org/10.1103/PhysRevLett.126.100402} {\bibfield  {journal}
  {\bibinfo  {journal} {Phys. Rev. Lett.}\ }\textbf {\bibinfo {volume} {126}},\
  \bibinfo {pages} {100402} (\bibinfo {year} {2021})}\BibitemShut {NoStop}%
\end{thebibliography}%
\end{document}